\documentclass[aps,twocolumn,showpacs,nofootinbib,floatsfix,amsmath,amssymb,prl]{revtex4}
\usepackage{amsmath}
\usepackage{graphicx}
\usepackage{bm}
\usepackage{color} 

\newcommand{\pb}{{\bf p}}

\newcommand{\ka}{\varkappa}
\newcommand{\Xb}{{\bf X}}
\newcommand{\xb}{{\bf x}}

\newcommand{\xbe}{{\bf x}_{\rm ext}}
\newcommand{\pbe}{{\bf p}_{\rm ext}}

\newcommand{\0}{^{(0)}}
\def\<{\lesssim}
\def\>{\gtrsim}

\begin{document}

\title{Disease extinction in the presence of non-Gaussian noise}

\author{Mark Dykman$^1$, Ira B. Schwartz$^{2}$,  Alexandra S. Landsman$^{2}$}

\affiliation{$^{1}$Department of Physics and Astronomy, Michigan State
University, East Lansing, MI 48824\\
$^2$US Naval Research Laboratory, Code
6792, Nonlinear Systems Dynamics Section, Plasma Physics Division,
Washington, DC 20375 }

\begin{abstract}
We investigate stochastic extinction in an epidemic model and the impact of random vaccinations in large populations. We show that, in the absence of vaccinations, the effective entropic barrier for extinction displays scaling with the distance to the bifurcation point, with an unusual critical exponent. Even a comparatively weak Poisson-distributed vaccination leads to an exponential increase in the extinction rate, with the exponent that strongly depends on the vaccination parameters.
\end{abstract}


\maketitle


Practically all diseases of interest exhibit randomness. Childhood
diseases \cite{AMbook,BolkerGrenfell93,Bolker93}, meningitis
\cite{Patz02b}, dengue fever and malaria \cite{Patz02a} are but a
few examples where incidence rates fluctuate. These fluctuations arise from fluctuations in population, epidemic parameters, and
intrinsically random contacts within the population
\cite{Rand-Wilson,BBS-PRL}. As diseases evolve in large
populations, there is the possibility of extinction and reintroduction of the
disease \cite{AnderssonB00,CummingsIHENUB04}. Extinction occurs where the
number of infectives become so small that there is insufficient transmission
to keep the disease in its endemic state
\cite{keeling04,verdaska05,bartlett49}. To gain  insight into disease
fade-out, one can think of the dynamics as coming from a nonlinear physical system
far from thermal equilibrium, with extinction resulting from a large
infrequent fluctuation.

A well-known model of population dynamics is the so-called SIS model where
only susceptibles (S) and infectives (I) are present
\cite{AMbook,jacquez93}. Here, in the absence of fluctuations the disease spread is characterized by the reproductive rate of
infection, $R_0$, which is defined so that an endemic state
exists along with the disease free equilibrium for $R_0>$1. The disease free state is unstable. However, in the presence of fluctuations, this state may be reached, albeit for a limited time
\cite{jacquez93,Herwaarden1995,west97,Nasell99,allen00,Elgart2004,citeulike:cumminhgs2005,Doering2005}

A major characteristic of fluctuation-induced extinction is the extinction
rate, or the reciprocal mean first time the number of infectives approaches
zero. It has been studied in the continuous limit using the Langevin approach with fluctuations induced by noise. A discrete birth-death SIS model was investigated recently and a comparison with the continuous model was performed by Doering {\it et al.} \cite{Doering2005}. The analysis referred to the one-variable model with detailed balance, which allows one to obtain an explicit solution. However, it does not reveal some generic features of the SIS model, including the scaling behavior of the extinction rate.

The goal of this letter is two-fold: (i) to show that the extinction rate scales with the control parameter $R_0$ and to find the
corresponding scaling exponent, and (ii) to study the effect of
vaccination on extinction rate. We will be interested in the important case where the vaccine schedule is a Poisson process. As we show, even comparatively weak vaccination can increase
the extinction rate exponentially strongly. This is a dynamical effect which happens, because an appropriate sequence of vaccine pulses can ``resonate" with the dynamics of the system followed during extinction.

We consider a model where susceptibles (S) are born at rate $\mu$,
both susceptibles and infectives (I) die at the same rate $\mu$, and infectives recover at rate $\ka$ and immediately become susceptible. If susceptibles contact infectives, they may become infected at rate $\beta$. Time-dependent vaccination reduces the number of susceptibles at rate $\xi(t)$ \cite{AMbook}. This rate will be assumed small, on average. The events of birth, death, and contact
happen at random. They are transitions between the states of the
system with different $S$ and $I$. Therefore the quantity of interest is the probability $\rho(S,I)$ to have given $S$ and $I$. It is given by the master equation
\begin{eqnarray}
\label{eq:master_master}
\dot \rho({\bf X})=\sum_r\left[W({\bf X-r; r})\rho({\bf X-r}) - W({\bf X; r})\rho({\bf X})\right]
\end{eqnarray}
Here, we introduced vector $\Xb = (X_1,X_2)$ with components
$X_1=S,X_2=I$ and vector ${\bf r}=(r_1,r_2)$ with components $r_1$ and $r_2$ showing, respectively, the increments in $S$ and $I$ in a single transition. The transition rates $W({\bf X, r})$ are
\begin{eqnarray}
\label{eq:rates}
&&W\bigl({\bf X}; (1, 0)\bigr)=N{}[\mu-\xi(t)],\quad W\bigl({\bf X}; (-1, 0)\bigr)=\mu X_1,\nonumber\\
&&W\bigl({\bf X}; (0, -1)\bigr)=\mu X_2,\quad W\bigl({\bf X}; (1, -1)\bigr)=\varkappa X_2,\nonumber\\
&&W\bigl({\bf X}; (-1, 1)\bigr)=\beta X_1X_2/N{},
\end{eqnarray}
where $N{}$ is the scaling factor which we set equal to the average
population, $N{}\gg 1$.

For large $S,I\propto N$ fluctuations of $S,I$ are small on average. If these fluctuations are disregarded, one arrives at the deterministic equations for the mean values of $S,I$
\begin{eqnarray}
\label{eq:mean_field}
&&\dot X_1= N{}[\mu-\xi(t)] -\mu X_1+ \varkappa X_2 - \beta X_1X_2/N{},\nonumber\\
&&\dot X_2 = -(\mu + \varkappa)X_2 + \beta X_1X_2/N{}.
\end{eqnarray}
These are standard equations of the SIS model. In the absence of vaccination,
$\xi(t)=0$, for $R_0=\beta/(\mu +\ka) > 1$ they have a stable endemic solution $\Xb_A=N\xb_A$ with $x_{1A}=R_0^{-1}, x_{2A}=1-R_0^{-1}$.  In addition, Eqs.~(\ref{eq:mean_field}) have an unstable stationary state (saddle point) $\Xb_{\cal S}=N\xb_{\cal S}$ with $x_{1{\cal S}}=1, x_{2{\cal S}}=0$. This state corresponds to extinction of infectives.

For $N\gg 1$ and for small vaccination rate the distribution $\rho(\Xb)$ has a peak at the stable state $\Xb_A$ with width $\propto N^{1/2}$. The probability of having a small number of infected, $X_2\ll X_{2A}$, is determined by the far tail of this peak. The tail can be obtained by seeking the solution of Eq.~(\ref{eq:master_master}) in the eikonal form,
\begin{eqnarray}
\label{eq:eikonal_general}
&&\rho({\bf X})=\exp[-N{}s({\bf x})], \qquad {\bf x}={\bf X}/N{} ,\nonumber\\
&&\rho({\bf X+r})\approx \rho({\bf X})\exp(-{\bf pr}), \qquad \pb =\partial_{\bf x}s.
\end{eqnarray}
For time-independent parameters $W$ this formulation was used in a number of papers \cite{Elgart2004,Doering2005,Kubo1973,Wentzell1976,Hu1987,Dykman1994d,Tretiakov2003}. However, in the present case it has special features, as shown below.

To leading order in $N{}^{-1}$, the equation for $s$ has a form of the Hamilton-Jacobi equation $\dot s=-H({\bf x},\partial_{\bf x}s;t)$, where $s$ is the effective action, and the effective Hamiltonian is
\begin{eqnarray}
\label{eq:Hamilton-Jacoby}
&& H({\bf x,p};t)=\sum_{\bf r}w({\bf x};{\bf r})\left(e^{\bf pr}-1\right),
\end{eqnarray}
with $w({\bf x;r})=N{}^{-1}W({\bf X}; {\bf r})$ being the transition rates per person. Action $s(\xb)$ can be found from classical trajectories of the auxiliary system with Hamiltonian $H$ that satisfy equations
\begin{eqnarray}
\label{eq:Hamilt_equations}
\dot\xb = \partial_{\bf p}H, \qquad \dot{\bf p}=-\partial_{\xb}H.
\end{eqnarray}

We start with the case where there is no vaccine, $\xi(t)=0$. In this case the transition rates $w=w\0$ and the Hamiltonian $H=H\0$ are independent of time. Of interest to us is the stationary distribution. The function $s=s{}\0$ is independent of time. It has the form
\cite{Wentzell1976,Hu1987,Dykman1994d,Tretiakov2003}
\begin{equation}
\label{eq:action_stationary}
s\0(\xb_f)=\int\nolimits_{-\infty}^{t_f} {\bf p}{}\,\dot\xb{} \,dt, \qquad  H{}\0(\xb{},{\bf p}{})=0.
\end{equation}
Here, the integral is calculated for a Hamiltonian trajectory
$\bigl(\xb{}(t{}),{\bf p}{}(t{})\bigr)$ that starts at $t\to -\infty$ at $\xb\to\xb_A, \pb \to {\bf 0}$ and arrives at time $t_f$ at a state $\xb_f$. This trajectory describes the most probable sequence of elementary events $\Xb\to\Xb+{\bf r}$ bringing the system to $\Xb_f=N\xb_f$. It provides the absolute minimum to $s\0(\xb_f)$ [$s\0(\xb_f)$ is independent of $t_f$]. The quantity $N{}s\0(\xb)$ is the entropic barrier for reaching $\Xb=N\xb$; it gives also the exponent in the expression for the mean first passage time for reaching $\Xb$ from the vicinity of the attractor $\Xb_A$ \cite{Matkowsky1984}.

The extinction rate is determined by $s\0$ for $x_2 \to 0$. It is
intuitively clear and can be shown from Eq.~(\ref{eq:Hamilton-Jacoby}) that the minimum of $s\0(\xb)$ over
$x_1$ for $x_2\to 0$ is reached at the saddle point $\xb_{\cal S}$ of the fluctuation-free motion (\ref{eq:mean_field}). Thus the entropic barrier for extinction is $Ns\0_{\rm ext}=Ns\0(\xb_{\cal S})$.

The Hamiltonian trajectory $\xbe{}(t),\pbe{}(t)$ that gives
$s\0(\xb_{\cal S})$ is the optimal extinction trajectory. One can show that it approaches $\xb_{\cal S}$ as $t\to \infty$. This is similar to other problems of an optimal trajectory leading from a deterministic stable state to a saddle point \cite{Dykman1994d,Maier1997}. However, in distinction from the more common situation, for $t\to\infty$ the momentum $\pbe{}$ does not go to zero. Instead $\pbe{}(t)\to \pb_{\cal S}$, with $\pb_{\cal S}=\bigl(0,-\log R_0\bigr)$. This is in spite of the fact that, along with $(\xb_{\cal S},\pb_{\cal S})$, the Hamiltonian $H\0$ has a ``standard" fixed point $(\xb_{\cal S},\pb={\bf 0})$.

Indeed, the optimal extinction trajectory should lie on the stable manifold of the fixed point with $\xb=\xb_{\cal S}$. The stable manifold of $(\xb_{\cal S},\pb ={\bf 0})$ lies on the plane $x_2=p_1=0$. The point $(\xb_A,\pb={\bf 0})$, and thus the optimal extinction trajectory do not lie on this plane. Therefore this trajectory may not go to $(\xb_{\cal S},\pb ={\bf 0})$. In contrast, it may go to the fixed point $(\xb_{\cal S},\pb_{\cal S})$, whose stable manifold is not confined to a plane in the $(\xb,\pb)$ space.

The situation where an auxiliary Hamiltonian system has two fixed points with the same $\xb_{\cal S}$ was first noticed for a system described by the Fokker-Planck equation with a singular at  $\xb_{\cal S}$ diffusion matrix \cite{Herwaarden1995}, and the ``right" point was chosen based on numerical simulations. This situation was also found for a system described by a one-variable master equation, where the Hamiltonian dynamics is integrable \cite{Elgart2004}; it occurs also in a two-variable susceptible-infected-recovered (SIR) model concurrently studied by Kamenev and Meerson \cite{Kamenev_private}.

Equations (\ref{eq:Hamilt_equations}) allow finding the extinction rate for any values of the parameters of the system. An explicit analytical solution in the absence of vaccination can be obtained near the saddle-node bifurcation point $R_0=1$ where states $\xb_A$ and $\xb_{\cal S}$ merge. For $\eta=\beta(R_0-1)/R_0 \ll 1$ we have $x_{2A}{}=\eta/\beta\ll 1$. The relaxation time of $x_2$ is $\eta^{-1}$. It is much longer than the relaxation time of $x_1{}$, which is $\mu^{-1}$, i.e., $x_2$ is a soft mode, and $x_1{}$ follows $x_2{}$ adiabatically.

In the adiabatic approximation we have in Hamiltonian equations (\ref{eq:Hamilt_equations}) $x_1=1-x_2, p_1=\beta x_2p_2/\mu$, while $x_2\ll 1, |p_2|\ll 1$. The equations for slow variables $x_2, p_2$ have a Hamiltonian form
\begin{eqnarray}
\label{eq:adiabatic}
  \dot x_2= \partial H^{\rm ad}/\partial p_2, \quad \dot p_2= -\partial H^{\rm ad}/\partial x_2
 \end{eqnarray}
with Hamiltonian $H^{\rm ad}=\eta x_2p_2-\beta x_2p_2(x_2-p_2)$. The Hamiltonian trajectory is
\begin{eqnarray}
\label{eq:explicit_adiabatic}
 p_{2}(t{})=x_{2}(t{})-\frac{\eta}{\beta},\quad x_{2}(t{}) = x_{2A}\left(1+e^{\eta(t{}-t{}_0)}\right)^{-1}.
\end{eqnarray}
From Eqs.~(\ref{eq:action_stationary}), (\ref{eq:explicit_adiabatic})
\begin{equation}
\label{eq:scaling_adiabat}
 s\0_{\rm ext}= \eta^{2}/2\beta^2=(R_0-1)^2/2R_0^2.
 \end{equation}
The entropic barrier for extinction $Ns\0_{\rm ext}$
(\ref{eq:scaling_adiabat}) scales with the distance to the
bifurcation point $\eta\propto R_0-1$ as $\eta^2$. This is in contrast to the standard scaling of the activation energy of escape near a saddle-node bifurcation point, where the critical exponent is $3/2$ \cite{Dykman1994d}, as has been seen in various dynamical systems close and far from thermal equilibrium. Such unusual scaling is related to $\pb_{\cal S}$ being nonzero. It emerges also in the SIR model \cite{Kamenev_private}.

\begin{figure}[h]
\includegraphics[width=2.4in]{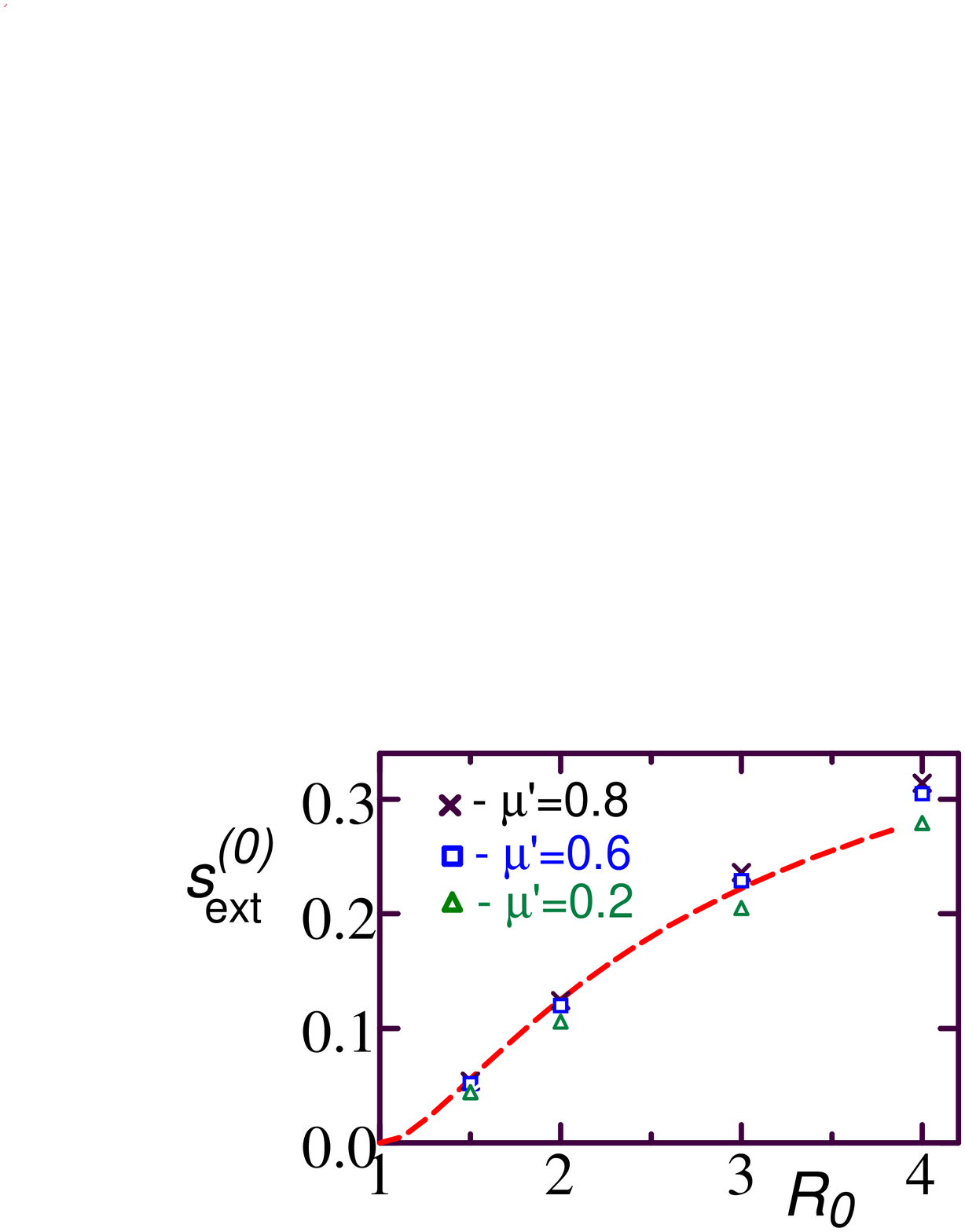}
\caption{The scaled barrier for extinction $s^{(0)}_{\rm ext}$ vs the reproductive rate of infection $R_0$. The dashed line shows asymptotic expression (\ref{eq:scaling_adiabat}). The data points for different $\mu'=\mu/(\mu+\varkappa)$ are obtained from a numerical solution of Eqs.~(\ref{eq:Hamilt_equations}), (\ref{eq:action_stationary}).}
\label{fig:s_ext}
\end{figure}

Generally, the scaled barrier $s^{(0)}_{\rm ext}$ depends on two parameters, $R_0$ and $\mu'=\mu/(\mu+\ka)$. In Fig.~\ref{fig:s_ext} asymptotic expression (\ref{eq:scaling_adiabat}) is compared with the results obtained from Eqs.~(\ref{eq:Hamilt_equations}), (\ref{eq:action_stationary}) for several $\mu'$. There is a reasonably good agreement even far from $R_0=1$. As $R_0$ increases the dependence of $s^{(0)}_{\rm ext}$ on $\mu'$ becomes more pronounced.

We now discuss the effect of vaccination. We are interested in the distribution $\langle \rho(\Xb)\rangle$
averaged over realizations of noise $\xi(t)$. If $\xi(t)$ is a stationary noise, $\langle \rho(\Xb)\rangle$ is stationary, too. The mean first time of reaching $\xb$ from the vicinity of the stable state is assumed to largely exceed the correlation time $t_{\rm corr}$ of $\xi(t)$. 

The full Hamiltonian of the system that determines $s$ can be written as $H=H\0+H^{(1)}$, with
\begin{equation}
\label{eq:H_1}
 H^{(1)}(\xb,\pb,t)= -\xi(t)h(\xb,\pb),\qquad h = \exp(p_1)-1.
\end{equation}
The term $H^{(1)}$ is small for weak noise $\xi(t)$. Because $s(\xb_f,t_f)$ provides a minimum to the integral over time of $\pb\,\dot\xb -H$, to first order in $\xi(t)$ we have \cite{LL_Mechanics2004}
\begin{equation}
\label{eq:s_1st_order}
 s(\xb_f,t_f)\approx s\0(\xb_f) +\int\nolimits_{-\infty}^{t_f}dt\xi(t)\chi_f(t).
\end{equation}
Here, $\chi_f(t)=h\bigl(\xb(t|\xb_f,t_f),\pb(t|\xb_f,t_f)\bigr)$; the function $\xb(t|\xb_f,t_f),\pb(t|\xb_ft_f)$ describes the Hamiltonian trajectory (\ref{eq:Hamilt_equations}) to $\xb_f$ calculated for $H=H^{(0)}$.

From Eq.~(\ref{eq:s_1st_order}), the logarithm of the distribution $\rho(\Xb_f,t_f)$ is linear in the force $\xi(t)$. The proportionality coefficient is $\propto \chi_f(t)$, and therefore we call $\chi_f(t)$ the logarithmic susceptibility, as for systems where fluctuations are induced by Gaussian noise \cite{Smelyanskiy1997c}; it is convenient to set $\chi_f(t)=0$ for $t>t_f$.

Equations (\ref{eq:eikonal_general}), (\ref{eq:s_1st_order}) give
\begin{eqnarray}
\label{eq:distribution_averaged}
 \langle \rho(\Xb)\rangle = A(\Xb)\rho\0(\Xb),\quad A(\Xb_f)=\tilde{\cal P}_{\xi}[iN{}\chi_f(t)],
 \end{eqnarray}
where $\rho\0$ is the distribution for $\xi(t)=0$ and $\tilde{\cal P}_{\xi}[\kappa(t)] = \left\langle\exp\left[i\int\kappa(t)\xi(t)dt\right]\right\rangle$ is the characteristic functional of $\xi(t)$. Because $N\gg 1$, an already weak noise $\xi(t)$ can significantly change the distribution $\langle \rho(\Xb)\rangle$, the factor $A$ can be exponentially large. 

To extend the analysis to the problem of extinction, we choose the final point to be $\xb_{\cal S}$. The corresponding logarithmic susceptibility $\chi_{\rm ext}(t)$ is determined by the unperturbed optimal extinction path $\left(\xbe(t),\pbe(t)\right)$. Since  $\xb_{\cal S}$ is approached along the optimal path asymptotically as $t\to \infty$, the action $s_{\rm ext}$ for reaching this point is given by Eq.~(\ref{eq:s_1st_order}) in which integration over time goes from $-\infty$ to $\infty$.  The noise-induced change of the extinction rate is then determined by the factor
\begin{eqnarray}
\label{eq:noise_extinct_factor}
 A_{\rm ext}=\tilde{\cal P}_{\xi}[iN\chi_{\rm ext}(t)], \quad
 \chi_{\rm ext}(t)=h\bigl(\xbe(t),\pbe(t)\bigr).
\end{eqnarray}
We call $A_{\rm ext}$ the noise-induced extinction factor.

The effect of noise on extinction can be illustrated using an important model where the noise is a Poisson process: a sequence of pulses $g(t-t_j)$ occurring at random times $t_j$. We assume that the pulses have a small amplitude and duration small compared to the relaxation time of the system and the reciprocal average pulse frequency $\nu^{-1}$ (the noise realizations of interest should satisfy the restriction $\xi(t) < \mu$). With account taken of the explicit form of the characteristic functional \cite{FeynmanQM},
\begin{eqnarray}
\label{eq:extinct_fact_Poisson}
 A_{\rm ext}=A_{\rm av}A_{\rm fl},\qquad A_{\rm av}=\exp\left[-\nu\int\nolimits_{-\infty}^{\infty}dt \,\kappa(t)\right],\\
 A_{\rm fl}=\exp\left[\nu\int\nolimits_{-\infty}^{\infty}dt\, \left(e^{-\kappa(t)}-1+ \kappa(t)\right)\right]\nonumber
\end{eqnarray}
with $\kappa(t)=N\int dt'\chi_{\rm ext}(t)g(t-t') \approx N\chi_{\rm ext}(t)\bar g$, where $\bar g=\int g(t)dt$. The factor $A_{\rm av}$ describes the effect of the average noise $\langle \xi(t)\rangle$, whereas the term $A_{\rm fl}$ describes the effect of the fluctuating part of $\xi(t)$ with zero mean.

The general expression (\ref{eq:extinct_fact_Poisson}) is simplified in the limiting cases. For weak noise, where $|\kappa(t)|\ll
1$, we have $A_{\rm fl}=\exp\left[\nu\int dt\,\kappa^2(t)/2\right]$,
implying $\log A_{\rm fl}\propto \bar g^2$. In the opposite limit,
$\max[-\kappa(t)]\gg 1$, we have $A_{\rm fl}=\exp\left[\nu (2\pi/\ddot
\kappa_m)^{1/2}\exp(-\kappa_m)\right]$, where $-\kappa_m\equiv
-\kappa(t_m)$ is the maximum of $-\kappa(t)$ and
$\ddot\kappa_m=\ddot\kappa(t_m)$. In this case $\log A_{\rm fl}$ is
exponential in the pulse intensity $\bar g$. Note that the results apply for $\log A_{\rm ext}\ll Ns^{(0)}$.

An explicit expression for $A_{\rm ext}$ can be obtained near the bifurcation point, $\eta\ll 1$. From Eq.~(\ref{eq:explicit_adiabatic}), here $\chi_{\rm ext}=\dot x_2/\mu$. This gives $A_{\rm av}= \exp[\nu\eta {\bar g} N/\beta\mu]$. The exponent in $A_{\rm av}$ linearly scales with the distance to the bifurcation point $\eta$ and the pulse intensity $\bar g$.

For pulse duration small compared to $\eta^{-1}$, the fluctuation part of the extinction factor $A_{\rm fl}$ is determined by the parameter $\sigma={\bar g}\eta^2N/\mu\beta$. If $\sigma\ll 1$, then $A_{\rm fl}\approx \exp[\nu\sigma^2/12\eta]$. In the opposite limit, $\sigma \gg 1$, we have $A_{\rm fl}\approx \exp[(4\nu/\eta)(\pi/\sigma)^{1/2}\exp(\sigma/4)]$. Here the dependence of $A_{\rm fl}$ on the vaccination pulse intensity $\bar g$ is double exponential.

\begin{figure}[h]
\includegraphics[width=2.4in]{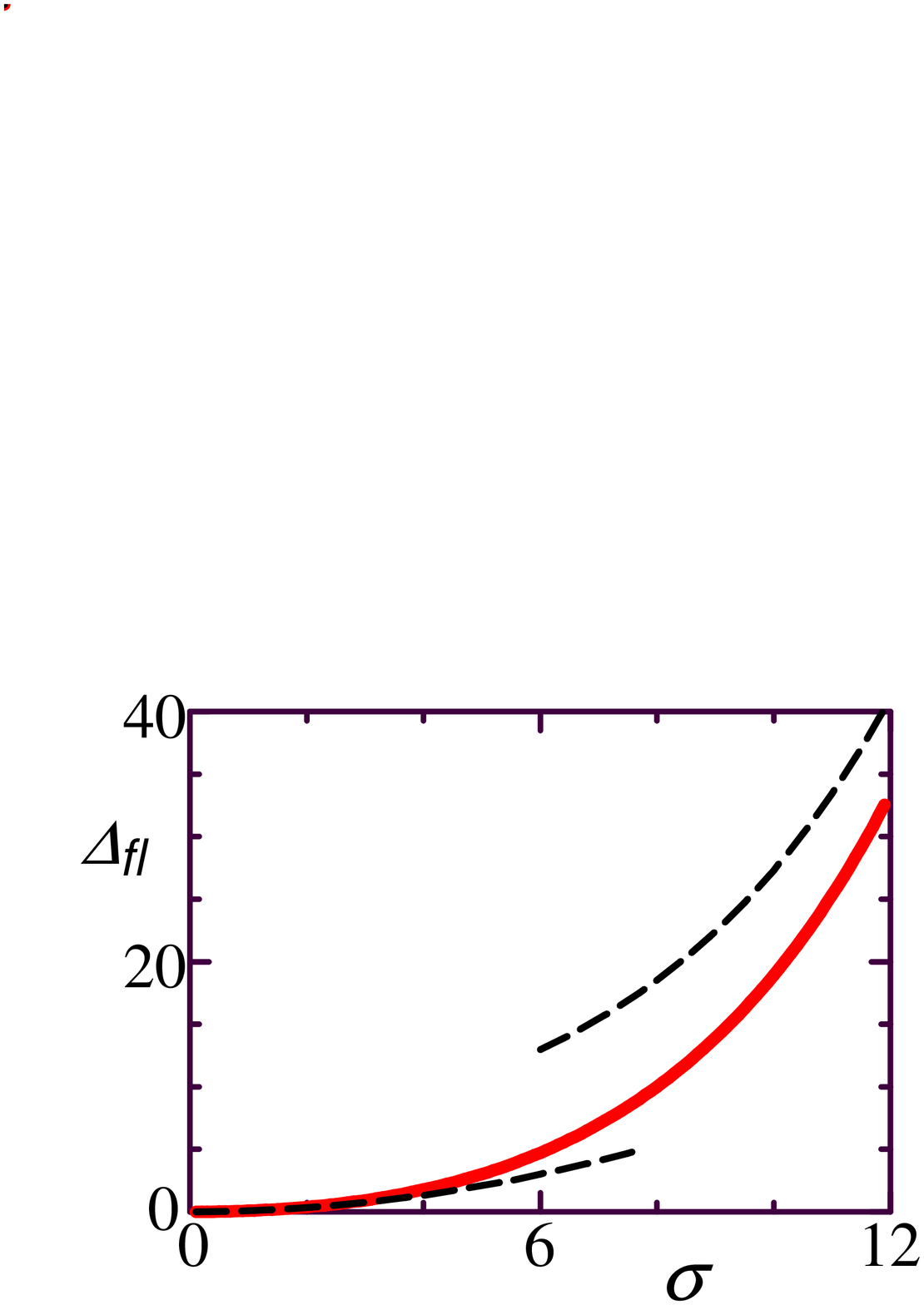}
\caption{The scaled fluctuation-induced change of the barrier for extinction, $\Delta_{\rm fl}=(\eta/\nu)\log A_{\rm fl}$, vs the scaled intensity of noise pulses $\sigma = {\bar g}\eta^2N/\mu\beta$ (solid line). The results refer to $\eta \ll 1$. The dashed lines show the asymptotic behavior of $\Delta_{\rm fl}$ for small and large $\sigma$.}
\label{fig:sis_poisson}
\end{figure}

The logarithm of  $A_{\rm fl}$ as a function of $\sigma$ is shown in Fig.~\ref{fig:sis_poisson}. The parabolic small-$\sigma$ asymptotics works reasonably well, numerically, for $\sigma \lesssim 3$, whereas the exponential large-$\sigma$ asymptotics is approached for $\sigma \gtrsim 15$.

The exponentially strong effect of random vaccine on disease extinction results from a dynamical cooperation between an outburst of noise of an appropriate temporal shape and the evolution of the system along the optimal path leading to extinction. One can think of the vaccine-induced change of the extinction barrier as a generalized work done by the corresponding noise realization along the optimal path. In this picture, a zero-mean noise should be expected to reduce the extinction barrier, and then the respective part of the noise-induced extinction factor $A_{\rm fl}$ should exceed 1. This is indeed the case for Poisson noise, as seen from Eq.~(\ref{eq:extinct_fact_Poisson}).

In summary, we have considered fluctuations in the full two-variable SIS model and found the rate of extinction of disease with and without vaccination. We have developed a general approach in which the problem is reduced to the analysis of dynamics of an auxiliary Hamiltonian system, with nontrivial boundary conditions. We show that the entropic barrier for extinction displays scaling dependence on the reproductive rate of infection $R_0$ for small $R_0-1$, with an unusual exponent. Even comparatively weak vaccination can exponentially strongly affect the extinction rate. A general expression that describes this effect for random vaccine in terms of its characteristic functional has been obtained. For a Poisson distributed vaccine, the change of the exponent of the extinction rate may itself depend on the vaccine strength exponentially.

The authors are grateful to A. Kamenev and B. Meerson for valuable discussions. This work was supported  by the  Army
Research Office and Office of Naval Research.


\begin{thebibliography}{34}
\expandafter\ifx\csname natexlab\endcsname\relax\def\natexlab#1{#1}\fi
\expandafter\ifx\csname bibnamefont\endcsname\relax
  \def\bibnamefont#1{#1}\fi
\expandafter\ifx\csname bibfnamefont\endcsname\relax
  \def\bibfnamefont#1{#1}\fi
\expandafter\ifx\csname citenamefont\endcsname\relax
  \def\citenamefont#1{#1}\fi
\expandafter\ifx\csname url\endcsname\relax
  \def\url#1{\texttt{#1}}\fi
\expandafter\ifx\csname urlprefix\endcsname\relax\def\urlprefix{URL }\fi
\providecommand{\bibinfo}[2]{#2}
\providecommand{\eprint}[2][]{\url{#2}}

\bibitem[{\citenamefont{Anderson and May}(1991)}]{AMbook}
\bibinfo{author}{\bibfnamefont{R.~M.} \bibnamefont{Anderson}} \bibnamefont{and}
  \bibinfo{author}{\bibfnamefont{R.~M.} \bibnamefont{May}},
  \emph{\bibinfo{title}{Infectious Diseases of Humans-Dynamics and Control}}
  (\bibinfo{publisher}{Oxford Science Publications}, \bibinfo{year}{1991}).

\bibitem[{\citenamefont{Bolker and Grenfell}(1993)}]{BolkerGrenfell93}
\bibinfo{author}{\bibfnamefont{B.~M.} \bibnamefont{Bolker}} \bibnamefont{and}
  \bibinfo{author}{\bibfnamefont{B.~T.} \bibnamefont{Grenfell}},
  \bibinfo{journal}{Proc. Roy. Soc. B} \textbf{\bibinfo{volume}{251}},
  \bibinfo{pages}{75} (\bibinfo{year}{1993}).

\bibitem[{\citenamefont{Bolker}(1993)}]{Bolker93}
\bibinfo{author}{\bibfnamefont{B.~M.} \bibnamefont{Bolker}},
  \bibinfo{journal}{IMA J. Math. Appl. Med.} \textbf{\bibinfo{volume}{10}},
  \bibinfo{pages}{83} (\bibinfo{year}{1993}).

\bibitem[{\citenamefont{Patz}(2002)}]{Patz02b}
\bibinfo{author}{\bibfnamefont{J.}~\bibnamefont{Patz}}, \bibinfo{journal}{Proc. Natl. Acad. Sci. USA}
  \textbf{\bibinfo{volume}{99}}, \bibinfo{pages}{12506} (\bibinfo{year}{2002}).

\bibitem[{\citenamefont{Patz et~al.}(2002)\citenamefont{Patz, Hulme,
  Rosenzweig, Mitchell, Goldberg, Githeko, Lele, {McMichael}, and {Le
  Sueur}}}]{Patz02a}
\bibinfo{author}{\bibfnamefont{J.}~\bibnamefont{Patz} {\it et al.}},
\bibinfo{journal}{Nature} \textbf{\bibinfo{volume}{420}},
  \bibinfo{pages}{627} (\bibinfo{year}{2002}).

\bibitem[{\citenamefont{Rand and Wilson}(1991)}]{Rand-Wilson}
\bibinfo{author}{\bibfnamefont{D.}~\bibnamefont{Rand}} \bibnamefont{and}
  \bibinfo{author}{\bibfnamefont{H.}~\bibnamefont{Wilson}},
  \bibinfo{journal}{Proc. Roy. Soc. B} \textbf{\bibinfo{volume}{246}},
  \bibinfo{pages}{179} (\bibinfo{year}{1991}).

\bibitem[{\citenamefont{Billings et~al.}(2002)\citenamefont{Billings, Bollt,
  and Schwartz}}]{BBS-PRL}
\bibinfo{author}{\bibfnamefont{L.}~\bibnamefont{Billings}},
  \bibinfo{author}{\bibfnamefont{E.}~\bibnamefont{Bollt}}, \bibnamefont{and}
  \bibinfo{author}{\bibfnamefont{I.}~\bibnamefont{Schwartz}},
  \bibinfo{journal}{Phys. Rev. Lett.} \textbf{\bibinfo{volume}{88}},
  \bibinfo{pages}{234101} (\bibinfo{year}{2002}).

\bibitem[{\citenamefont{Andersson and Britton}(2000)}]{AnderssonB00}
\bibinfo{author}{\bibfnamefont{H.}~\bibnamefont{Andersson}} \bibnamefont{and}
  \bibinfo{author}{\bibfnamefont{T.}~\bibnamefont{Britton}},
  \bibinfo{journal}{J. Math. Biology}
  \textbf{\bibinfo{volume}{41}}, \bibinfo{pages}{559} (\bibinfo{year}{2000}).

\bibitem[{\citenamefont{Cummings et~al.}(2004)\citenamefont{Cummings, Irizarry,
  Huang, Endy, Nisalak, Ungchusak, and Burke}}]{CummingsIHENUB04}
\bibinfo{author}{\bibfnamefont{D.~A.~T.} \bibnamefont{Cummings} {\it et al.}},
  \bibinfo{journal}{Nature} \textbf{\bibinfo{volume}{427}},
  \bibinfo{pages}{344} (\bibinfo{year}{2004}).

\bibitem[{\citenamefont{Keeling}(2004)}]{keeling04}
\bibinfo{author}{\bibfnamefont{M.~J.} \bibnamefont{Keeling}},
  \emph{\bibinfo{title}{Ecology, Genetics, and Evolution.}}
  (\bibinfo{publisher}{Elsevier}, \bibinfo{address}{New York},
  \bibinfo{year}{2004}).

\bibitem[{\citenamefont{Verdasca and et~al}(2005)}]{verdaska05}
\bibinfo{author}{\bibfnamefont{J.}~\bibnamefont{Verdasca}} \bibnamefont{and}
  \bibinfo{author}{\bibnamefont{et~al}}, \bibinfo{journal}{J. Theor. Biology}
  \textbf{\bibinfo{volume}{233}}, \bibinfo{pages}{553} (\bibinfo{year}{2005}).

\bibitem[{\citenamefont{Bartlett}(1949)}]{bartlett49}
\bibinfo{author}{\bibfnamefont{M.~S.} \bibnamefont{Bartlett}},
  \bibinfo{journal}{J. Roy. Stat. Soc. B} \textbf{\bibinfo{volume}{11}},
  \bibinfo{pages}{211} (\bibinfo{year}{1949}).

\bibitem[{\citenamefont{Jacquez and Simon}(1993)}]{jacquez93}
\bibinfo{author}{\bibfnamefont{J.~A.} \bibnamefont{Jacquez}} \bibnamefont{and}
  \bibinfo{author}{\bibfnamefont{C.~P.} \bibnamefont{Simon}},
  \bibinfo{journal}{Math. Biosci.} \textbf{\bibinfo{volume}{117}},
  \bibinfo{pages}{77} (\bibinfo{year}{1993}).

\bibitem[{\citenamefont{van Herwaarden and Grasman}(1995)}]{Herwaarden1995}
\bibinfo{author}{\bibfnamefont{O.~A.} \bibnamefont{van Herwaarden}}
  \bibnamefont{and} \bibinfo{author}{\bibfnamefont{J.}~\bibnamefont{Grasman}},
  \bibinfo{journal}{J. Math. Biology} \textbf{\bibinfo{volume}{33}},
  \bibinfo{pages}{581} (\bibinfo{year}{1995}).

\bibitem[{\citenamefont{West and Thompson}(1997)}]{west97}
\bibinfo{author}{\bibfnamefont{R.~W.} \bibnamefont{West}} \bibnamefont{and}
  \bibinfo{author}{\bibfnamefont{J.~R.} \bibnamefont{Thompson}},
  \bibinfo{journal}{Math. Biosci.} \textbf{\bibinfo{volume}{141}},
  \bibinfo{pages}{29} (\bibinfo{year}{1997}).

\bibitem[{\citenamefont{Nasell}(1999)}]{Nasell99}
\bibinfo{author}{\bibfnamefont{I.}~\bibnamefont{Nasell}},
  \bibinfo{journal}{Math. Biosci.} \textbf{\bibinfo{volume}{156}},
  \bibinfo{pages}{21} (\bibinfo{year}{1999}).

\bibitem[{\citenamefont{Allen and Burgin}(2000)}]{allen00}
\bibinfo{author}{\bibfnamefont{L.}~\bibnamefont{Allen}} \bibnamefont{and}
  \bibinfo{author}{\bibfnamefont{A.~M.} \bibnamefont{Burgin}},
  \bibinfo{journal}{Math. Biosci.} \textbf{\bibinfo{volume}{163}},
  \bibinfo{pages}{1} (\bibinfo{year}{2000}).

\bibitem[{\citenamefont{Elgart and Kamenev}(2004)}]{Elgart2004}
\bibinfo{author}{\bibfnamefont{V.}~\bibnamefont{Elgart}} \bibnamefont{and}
  \bibinfo{author}{\bibfnamefont{A.}~\bibnamefont{Kamenev}},
  \bibinfo{journal}{Phys. Rev. E} \textbf{\bibinfo{volume}{70}},
  \bibinfo{pages}{041106} (\bibinfo{year}{2004}).

\bibitem[{\citenamefont{Cummings et~al.}(2005)}]{citeulike:cumminhgs2005}
\bibinfo{author}{\bibfnamefont{D.~A.} \bibnamefont{Cummings}} \bibnamefont{and}
  \bibinfo{author}{\bibnamefont{et~al.}},
\bibinfo{journal}{Proc. Natl. Acad. Sci. USA} \textbf{\bibinfo{volume}{102}}, \bibinfo{pages}{15259}
  (\bibinfo{year}{2005}).


\bibitem[{\citenamefont{Doering et~al.}(2005)\citenamefont{Doering, Sargsyan,
  and Sander}}]{Doering2005}
\bibinfo{author}{\bibfnamefont{C.~R.} \bibnamefont{Doering}},
  \bibinfo{author}{\bibfnamefont{K.~V.} \bibnamefont{Sargsyan}},
  \bibnamefont{and} \bibinfo{author}{\bibfnamefont{L.~M.}
  \bibnamefont{Sander}}, \bibinfo{journal}{Multiscale Mod. Sim.}
  \textbf{\bibinfo{volume}{3}}, \bibinfo{pages}{283} (\bibinfo{year}{2005}).

\bibitem[{\citenamefont{Kubo et~al.}(1973)\citenamefont{Kubo, Matsuo, and
  Kitahara}}]{Kubo1973}
\bibinfo{author}{\bibfnamefont{R.}~\bibnamefont{Kubo}},
  \bibinfo{author}{\bibfnamefont{K.}~\bibnamefont{Matsuo}}, \bibnamefont{and}
  \bibinfo{author}{\bibfnamefont{K.}~\bibnamefont{Kitahara}},
  \bibinfo{journal}{J. Stat. Phys.} \textbf{\bibinfo{volume}{9}},
  \bibinfo{pages}{51} (\bibinfo{year}{1973}).

\bibitem[{\citenamefont{Ventcel'}(1976)}]{Wentzell1976}
\bibinfo{author}{\bibfnamefont{A.~D.} \bibnamefont{Ventcel'}},
  \bibinfo{journal}{Teor. Verojatnost. Primenen.}
  \textbf{\bibinfo{volume}{21}}, \bibinfo{pages}{235} (\bibinfo{year}{1976}).

\bibitem[{\citenamefont{Hu}(1987)}]{Hu1987}
\bibinfo{author}{\bibfnamefont{G.}~\bibnamefont{Hu}}, \bibinfo{journal}{Phys.
  Rev. A} \textbf{\bibinfo{volume}{36}}, \bibinfo{pages}{5782}
  (\bibinfo{year}{1987}).

\bibitem[{\citenamefont{Dykman et~al.}(1994)\citenamefont{Dykman, Mori, Ross,
  and Hunt}}]{Dykman1994d}
\bibinfo{author}{\bibfnamefont{M.~I.} \bibnamefont{Dykman}},
  \bibinfo{author}{\bibfnamefont{E.}~\bibnamefont{Mori}},
 \bibinfo{author}{\bibfnamefont{J.}~\bibnamefont{Ross}}, \bibnamefont{and}
  \bibinfo{author}{\bibfnamefont{P.~M.} \bibnamefont{Hunt}},
  \bibinfo{journal}{J. Chem. Phys.} \textbf{\bibinfo{volume}{100}},
  \bibinfo{pages}{5735} (\bibinfo{year}{1994}).

\bibitem[{\citenamefont{Tretiakov et~al.}(2003)\citenamefont{Tretiakov,
  Gramespacher, and Matveev}}]{Tretiakov2003}
\bibinfo{author}{\bibfnamefont{O.~A.} \bibnamefont{Tretiakov}},
  \bibinfo{author}{\bibfnamefont{T.}~\bibnamefont{Gramespacher}},
  \bibnamefont{and} \bibinfo{author}{\bibfnamefont{K.~A.}
  \bibnamefont{Matveev}}, \bibinfo{journal}{Phys. Rev. B}
  \textbf{\bibinfo{volume}{67}}, \bibinfo{pages}{073303}
  (\bibinfo{year}{2003}).

\bibitem[{\citenamefont{Matkowsky et~al.}(1984)\citenamefont{Matkowsky, Schuss,
  Knessl, Tier, and Mangel}}]{Matkowsky1984}
\bibinfo{author}{\bibfnamefont{B.}~\bibnamefont{Matkowsky et al.}},
  \bibinfo{journal}{Phys. Rev. A} \textbf{\bibinfo{volume}{29}},
  \bibinfo{pages}{3359 } (\bibinfo{year}{1984}).

\bibitem[{\citenamefont{Maier and Stein}(1997)}]{Maier1997}
\bibinfo{author}{\bibfnamefont{R.~S.} \bibnamefont{Maier}} \bibnamefont{and}
  \bibinfo{author}{\bibfnamefont{D.~L.} \bibnamefont{Stein}},
  \bibinfo{journal}{SIAM J. Appl. Math.} \textbf{\bibinfo{volume}{57}},
  \bibinfo{pages}{752} (\bibinfo{year}{1997}).

\bibitem[{Kam()}]{Kamenev_private}
\bibinfo{howpublished}{A. Kamenev and B. Meerson, in preparation}.

\bibitem[{\citenamefont{Landau and Lifshitz}(2004)}]{LL_Mechanics2004}
\bibinfo{author}{\bibfnamefont{L.~D.} \bibnamefont{Landau}} \bibnamefont{and}
  \bibinfo{author}{\bibfnamefont{E.~M.} \bibnamefont{Lifshitz}},
  \emph{\bibinfo{title}{Mechanics}} (\bibinfo{publisher}{Elsevier, Amsterdam},
  \bibinfo{year}{2004}), \bibinfo{edition}{3rd} ed.

\bibitem[{\citenamefont{Smelyanskiy et~al.}(1997)\citenamefont{Smelyanskiy,
  Dykman, Rabitz, and Vugmeister}}]{Smelyanskiy1997c}
\bibinfo{author}{\bibfnamefont{V.~N.} \bibnamefont{Smelyanskiy} {\it et al.}},
  \bibinfo{journal}{Phys. Rev. Lett.} \textbf{\bibinfo{volume}{79}},
  \bibinfo{pages}{3113} (\bibinfo{year}{1997}).

\bibitem[{\citenamefont{Feynman and Hibbs}(1965)}]{FeynmanQM}
\bibinfo{author}{\bibfnamefont{R.~P.} \bibnamefont{Feynman}} \bibnamefont{and}
  \bibinfo{author}{\bibfnamefont{A.~R.} \bibnamefont{Hibbs}},
  \emph{\bibinfo{title}{Quantum Mechanics and Path Integrals}}
  (\bibinfo{publisher}{McGraw-Hill}, \bibinfo{address}{New-York},
  \bibinfo{year}{1965}).

\end{thebibliography}

\end{document}